\documentclass[epj]{svjour}
\usepackage{graphics}

\usepackage{amssymb}
\usepackage{epsfig}
\newcommand{\be}{\begin{equation}}
\newcommand{\ee}{\end{equation}}
\newcommand{\ba}{\begin{eqnarray}}
\newcommand{\ea}{\end{eqnarray}}
\newcommand{\msbar}{\overline{\mbox{\scriptsize MS}}}

\begin{document}
\title{A lattice perspective of kaon phenomenology
\thanks{Invited talk at the Int. Europhysics Conference on High 
Energy Physics HEP2003, July 2003, Aachen, Germany.}}
\author{Leonardo Giusti
\thanks{On leave from Centre de Physique Th\'eorique, CNRS Luminy, Case 907, 
 F-13288 Marseille, France.}
}                     
%
%
\institute{CERN, Theory Division, CH-1211 Geneva, Switzerland}
%
%
\abstract{I review recent lattice computations 
of the matrix element relevant for $K^0-\overline{K}^0$ mixing and discuss 
the advantages of fermions with an exact chiral symmetry to compute $K\rightarrow \pi\pi$ amplitudes.
} 
\maketitle

\section{Introduction}\label{intro}
Kaon decays are sources of key informations 
to test the flavour structure of the Standard Model and 
to search for new physics. Recently 
the experimental results for indirect CP violation 
in $K\rightarrow\pi\pi$ decays and those from B 
decays and oscillations were compared to
test the Cabibbo--Kobayashi--Maskawa picture of flavour mixing 
and CP violation \cite{Ciuchini:2000de,Hocker:2001xe,Battaglia:2003in}. 
In these analyses the experimental result for $\varepsilon$ is connected to 
the fundamental parameters of the underlying electroweak theory
through the lattice determination of the relevant hadronic 
matrix element of the $\Delta S =2$ effective Hamiltonian.
Its bag parameter $\hat B_K$ has been computed in the 
quenched approximation over the past decade with a 
remarkable accuracy. A summary estimate of $\hat B_K$ is given 
in the first part of this talk, followed by a review of 
recent computations.

Quantitative tests of the same quality have not been 
possible so far in the $\Delta S =1$ sector for 
the so-called $\Delta I=1/2$ rule in $K\rightarrow\pi\pi$ 
decays or for the direct CP violation parameter 
$\varepsilon'/\varepsilon$. In this sector a more complicated blend 
of ultraviolet and infrared effects renders 
the determination of the relevant matrix elements 
very challenging and prevented so far reliable 
computations with standard regularizations.
In the ultraviolet, power-divergent subtractions can be needed
to construct the renormalized operators that enter 
the effective Hamiltonian 
\cite{Bernard:wf,Maiani:1986db,Dawson:1997ic,Capitani:2000bm}. 
In the infrared, the continuation of the theory to Euclidean space-time 
and the use of finite volumes in numerical simulations generate a 
non-simple relation between the physical amplitudes and those computed 
on the lattice \cite{Maiani:ca,Lellouch:2000pv,Lin:2001ek}. 
The use of lattice fermion discretizations which preserve an 
exact chiral symmetry at finite lattice spacing
greatly simplifies some of these computations 
and allows one to attack them as shown in the second 
part of this talk\footnote{For lack of space
only a selection of topics are reviewed here, chosen following 
my personal taste and competence. Please see  
Ref.~\cite{Isidori:2001nd} for an up-to-date report
on rare kaon decays and Ref.~\cite{lattice2003}
for a broader overview on the lattice activity 
in kaon physics.}. 

Before entering into details, it is important to stress 
that lattice QCD allows one to perform non-perturbative 
computations of masses and matrix elements from first principles,
with control, at least in principle, over all sources of 
systematic errors. These uncertainties can be systematically 
reduced by exploiting the properties of the underlying quantum field 
theory and/or by using more powerful computers, but 
without the introduction of any model-dependent assumption or parameter. 
Examples are given by the $O(a)$-improved actions 
and operators and by the non-perturbative techniques 
used to renormalize bare matrix elements. 
Up to now many computations are still performed in 
the so-called quenched approximation, i.e. dropping the vacuum polarization 
effects in the Monte Carlo simulations. For some of them this represents 
the only source of systematic uncertainty not under control.

\vspace{-0.5cm}
\section{K$\rightarrow \pi\pi$ decays}
Non-leptonic $K\rightarrow \pi\pi$ amplitudes can be parametrized as 
\ba 
T[K^+\rightarrow\pi^+\pi^0] & = & \sqrt{3\over 2} A_2 e^{i\delta_2}\\
T[K^0\rightarrow\pi^+\pi^-] & = & \sqrt{2\over 3} A_0 e^{i\delta_0}+ \sqrt{1\over
3} A_2 e^{i\delta_2}\nonumber\\
T[K^0\rightarrow\pi^0\pi^0] & = & \sqrt{2\over 3} A_0 e^{i\delta_0}-2\sqrt{1\over
3} A_2 e^{i\delta_2}\, ,\nonumber
\ea
where $\delta_{I}$ and $A_I$ are the $\pi\pi$ phase shifts and 
the isospin amplitudes for $I=0,2$. Direct and indirect CP violations are parametrized by  
\be
\varepsilon' = \frac{1}{\sqrt{2}} {\rm e}^{i\Phi}
\frac{{\rm Re}A_{2}}{{\rm Re}A_{0}}
\left(\frac{{\rm Im}A_{2}}{{\rm Re}A_{2}}-\frac{{\rm Im}A_{0}} {{\rm Re}A_{0}}\right)
\ee
and 
\be
\varepsilon = \frac{T[K_L\rightarrow (\pi\pi)_{0}]}
                  {T[K_S\rightarrow (\pi\pi)_{0}]} 
\ee
respectively. Experimental results reveal $\Phi=\pi/2 + \delta_2 - \delta_0\thickapprox \pi/4$,
a $\Delta I=1/2$ selection rule $|A_0/A_2| \simeq  22.2$,
the presence of  
direct and indirect CP violation in nature
\ba
\mbox{Re}(\varepsilon'/\varepsilon) & = &  (16.6 \pm 1.6)\times 10^{-4} \;\; \;\;\;\;\;
\mbox{\cite{Batley:2002gn,Alavi-Harati:2002ye}} \nonumber\\
 | \varepsilon |\;\;\; & = & (2.282 \pm 0.017)\times 10^{-3}\;\; \mbox{\cite{pdg2002}}\; .
\ea
The large value of  $|A_0/A_2|$ and that of $\varepsilon^\prime/\varepsilon$ 
can be explained within the Standard Model only if the strong 
interactions crucially affect these non-leptonic weak transitions 
(see \cite{Ciuchini:1999xi,Bertolini:2002rj,deRafael:2002tj} for recent reviews). 

\vspace{-0.5cm}
\subsection{$K^0$--$\bar{K}^0$ mixing: $\varepsilon$}
\begin{figure}
\vskip 0.75cm
\resizebox{0.50\textwidth}{!}{%
\epsfig{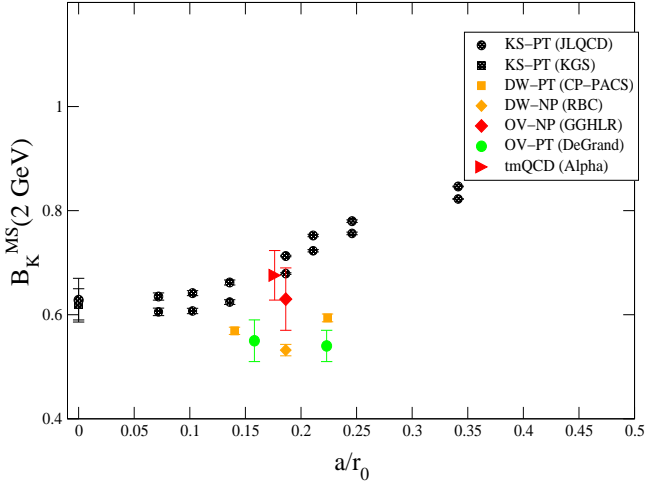}
}
\caption{$B^{\msbar}_K(2\, {\rm GeV})$ versus the lattice spacing 
in units of the Sommer scale $r_0$ \cite{Guagnelli:1998ud}
from the reference computation, black circles \cite{Aoki:1997nr} 
(see also also black squares \cite{Kilcup:1997ye}), and from recent 
determinations: red diamonds \cite{Garron:2003cb}, green circles \cite{DeGrand:2003in}, 
red triangle \cite{Dimopoulos:2003kc},  
orange diamonds \cite{Blum:2001xb}, orange squares \cite{AliKhan:2001wr}.}
\label{fig:bk}       
\end{figure}

By using the OPE, the 
$\Delta S =2$ effective Hamiltonian is given by 
\be
H^{\Delta S=2}_{\rm eff} = \frac{G^2_F M^2_W}{4\pi^2}
 {\cal C}_1(\mu) \, \widehat{ \cal O}_1 (\mu) + {\rm h. c.}\; ,
\ee
where the expression of the Wilson coefficient $ {\cal C}_1(\mu)$ 
as a function of the fundamental parameters of the 
underlying theory is known at NLO \cite{Herrlich:1995hh} and the corresponding
bare four-fermion operator is given by
\be
{\cal O}_{1} = (\bar s \gamma_\mu P_- d) (\bar s \gamma_\mu P_- d )\, ,
\qquad P_\pm = \frac{1\pm \gamma_5}{2}\, .
\ee
Its matrix element between $\overline{K}^0$ and $K^0$
\be
\langle \overline{K}^0| \widehat {\cal O}_1 |K^0\rangle \equiv
\frac{4}{3}  F_K^2 M_K^2 \widehat B_K
\ee
encodes the long-distance QCD contributions to $\varepsilon$.
So far the most precise and careful computation 
of $\widehat B_K(\mu)$ in the quenched approximation
has been carried out by JLQCD \cite{Aoki:1997nr}
(see also \cite{Kilcup:1997ye}).
They used staggered fermions, one-loop perturbation theory 
to renormalize the operator and degenerate down 
and strange quark masses. Their results  
at fixed lattice spacings are reported in Fig.~\ref{fig:bk}
together with their continuum extrapolated value 
\ba\label{eq:bk}
\widehat B_K^{\msbar}(2 \mbox{GeV}) & = & 0.63 \pm 0.04 \\
\hat B_K & = & 0.86 \pm 0.06 \; ,
\ea
where $\hat B_K$ is the renormalization group-invariant 
bag parameter.
The final error of $0.04$ is much larger 
than the statistical ones at fixed lattice spacing.
The amplification is due to the fact that with these 
fermions $O(a^2)$ effects in the matrix element 
are found to be large, and 
a consistent continuum limit of the results can be obtained
only if  also $O(\alpha_s^2)$ corrections are allowed 
in the extrapolation fit. By using chiral perturbation theory,
Sharpe estimates a $5 \%$ systematics due to the 
degeneracy of the quark masses \cite{Sharpe:1998hh}. 
At present the systematic error due to the quenching 
approximation is not under control. A crude estimate
based on quenched chiral perturbation theory and 
partially quenched simulations suggests to include 
a further $15 \% $ error in Eq.~(\ref{eq:bk}) \cite{Sharpe:1998hh}.
The error on $\hat B_K$ can be put under control and 
significantly reduced only with full QCD simulations 
with realistic light-quark masses. Despite a great effort in 
the last few years in the lattice community 
\cite{Gottlieb:2003bt,Jansen:2003nt}, the simulation
of dynamical quarks is still 
very challenging and a breakthrough in numerical algorithms 
may still be needed to reach interesting light-quark masses.\\
In the last year several collaborations have been computing 
$\hat B_K$ in quenched QCD by using sophisticated regularizations, 
non-perturbative renormalization techniques and allowing 
for $m_s\neq m_d$
\cite{Garron:2003cb,DeGrand:2003in,Dimopoulos:2003kc}. 
The aim is to check the staggered result and to pin down 
the systematic error within the quenched approximation. 
Even if a significant comparison can be performed 
only when all systematics (within the quenching 
approximation) will be under control also in these computations, 
it is interesting to analyse and compare the results to appreciate
their potentiality.\\  
Two collaborations~\cite{Garron:2003cb,DeGrand:2003in} computed $\hat B_K$
with Ginsparg--Wilson fermions 
\cite{Ginsparg:1982bj,Kaplan:1992bt,Neuberger,PerfectA}
for the first time. 
These regularizations simultaneously preserve chiral and flavour 
symmetries at finite lattice spacing \cite{Luscher:1998pq}.
As a consequence ${\cal O}_{1}$ renormalizes multiplicatively 
and its matrix elements are $O(a)$-improved.
The plain Neuberger action \cite{Neuberger} has been used in 
Ref.~\cite{Garron:2003cb} for a lattice of linear extension 
$L\simeq 1.5$~fm, a spacing $a\simeq0.09$~fm
and for degenerate light-quark masses. The 
RI/MOM non-perturbative renormalization procedure
has been implemented to compute the logarithmic divergent 
renormalization constant. The result obtained is reported 
in Fig.~\ref{fig:bk}. Even with a large error, it is in very good 
agreement with the continuum extrapolated staggered result,
suggesting that also for this matrix element discretization 
effects may be mild with Neuberger's fermions.
A NNC--HYP overlap fermion action 
has been implemented in Ref.~\cite{DeGrand:2003in} for lattices 
of linear extension $L\simeq 1.3$~fm, $a\simeq 0.08,0.12$~fm 
and for degenerate light-quark masses. 
The results are lower but still compatible 
within two sigmas with the continuum extrapolated 
staggered one and are in the same ballpark as those 
obtained with domain-wall fermions with a finite fifth 
dimension \cite{Blum:2001xb,AliKhan:2001wr}.
Even if both studies require more work to properly assess the 
magnitude of the various systematic errors, they confirmed
that Ginsparg--Wilson fermions can be very effective in 
computing phenomenological quantities despite 
their numerical cost.\\
In the last few years the ALPHA collaboration has been 
computing $\hat B_K$ with the goal of reaching a very 
precise determination of $\hat B_K$ within the quenched 
approximation. They are using twisted-mass fermions, which 
break flavour symmetry and violate parity but allow for 
a multiplicatively renormalization of ${\cal O}_{1}$.
A non-perturbative determination of the renormalization 
constant and its running has been completed, while the 
computation of the matrix element is still under way. 
A preliminary result has been reported in \cite{Dimopoulos:2003kc}
and is shown in Fig.~\ref{fig:bk}. Also in this case the result
obtained at $a\simeq 0.09$ is in very good 
agreement with the continuum extrapolated staggered result.

\vspace{-0.5cm}
\subsection{The $\Delta I=1/2$ rule}
By using the operator product expansion (OPE), the CP-conserving 
$\Delta S =1$ effective Hamiltonian above the charm threshold is 
given by \cite{Weinberg:1973hx,Gaillard:1974nj,Altarelli:1974ex}
\[
H_{\rm eff}^{\Delta S =1} =  
\frac {G_F} {\sqrt{2}}
\Bigl[ C_{+}(\mu) \widehat {\cal O}_{+}(\mu) + 
       C_{-}(\mu) \widehat {\cal O}_{-}(\mu) \Bigr]\; ,
\]
where the Wilson coefficients $C_{\pm}(\mu)$ 
are known at the NLO \cite{Buras:1993dy,Ciuchini:1993vr}
and the bare operators are 
\ba
{\cal O}_{\pm} & = & \Bigl[({\bar s}^{a} \gamma_\mu P_-  u^{b} )
        ({\bar u}^{b} \gamma_\mu P_-  d^{a}) \\
& \pm &  
({\bar s} \gamma_\mu P_-  u)
        ({\bar u} \gamma_\mu P_-  d)\Bigr]
                -  (u \rightarrow c)\; .\nonumber
\ea
The contributions that arise when the top quark is integrated out
are heavily suppressed by CKM factors  
and can be neglected. ${\cal O}_{\pm}$ belong 
to different chiral multiplets and are CPS-even. 
In correlation functions at non-zero physical distance, ${\cal O}_{\pm}$ cannot
mix between themselves or with other four-fermion operators
if the regularization preserves chiral symmetry,  
but only with the dimension-six operator 
\cite{Maiani:1986db,Capitani:2000bm}
\be
{\cal Q}_m = (m^2_u -m^2_c)\Bigl[m_d (\bar s P_+ d) + 
m_s (\bar s P_- d)\Bigr]\; .
\ee
The renormalized operators are
\be
\widehat {\cal O}_{\pm}(\mu) =  Z_{\pm}(\mu) 
\Bigl[ {\cal O}_{\pm} + b_{\pm}^m {\cal Q}_{m}\Bigr]\; , \;\;
\ee
where $Z_{\pm}(\mu)$ are logarithmic-divergent 
renormalization constants and $b_{\pm}^m$ are 
suppressed by a factor  $\alpha_s$.
{\it No power-divergent subtractions are needed}
to renormalize ${\cal O}_{\pm}$ when fermions with an exact 
chiral symmetry are used \cite{Capitani:2000bm}.

For $m_s \neq m_d$,
\ba
{\cal Q}_m & = & (m^2_u -m^2_c) 
\partial_\mu\Bigl[\frac{m_d+m_s}{m_s-m_d}\, {\cal V}_\mu^{sd}\\
& + & 
\frac{m_d-m_s}{m_s+m_d}\, {\cal A}_\mu^{sd} \Bigr]\nonumber
\ea
and it does not contribute to matrix elements that 
preserve four-momentum \cite{Bernard:wf,Maiani:1986db}.

If the charm is integrated out, not only potentially large
contributions of $O(\mu^2/m_c^2)$ are neglected,
but ultraviolet power divergences can arise
in the renormalization pattern of the relevant 
four-fermion operators. In this case the 
$\Delta S =1 $ effective Hamiltonian can be written as
\be\label{eq:nocharm}
H_{\rm eff}^{\Delta S =1} =  \displaystyle \frac {G_F} {\sqrt{2}} 
\sum_{i=1}^{10} C_i(\mu) \widehat {\cal Q}_i(\mu) \; .
\ee 
The so-called QCD-penguin operators are 
\ba
{\cal Q}_{3,5} &=& ({\bar s} \gamma_\mu P_- d)
    \sum_{q=u,d,s}({\bar q} \gamma_\mu P_\mp q) \\
{\cal Q}_{4,6} &=& ({\bar s}^{a} \gamma_\mu P_- d^{b})
    \sum_{q=u,d,s}({\bar q}^{b} \gamma_\mu P_\mp  q^{a})
\ea
(see Refs.~\cite{Ciuchini:1999xi,Bertolini:2002rj}
for definitions of the other operators).
At non-zero physical distance, mixing with two lower-dimensional operators 
\ba
{\cal Q}_p & = & m_d (\bar s P_+  d) + m_s (\bar s P_-  d)\\
{\cal Q}_\sigma & = & m_d (\bar s F_{\mu\nu} \sigma_{\mu\nu}P_+  d)
+ m_s (\bar s F_{\mu\nu} \sigma_{\mu\nu}P_-  d)\nonumber
\ea
can occur and power-divergent subtractions are needed even 
with Ginsparg--Wilson fermions.

With Wilson fermions, 
only CPS and flavour symmetry can be used to determine the 
renormalization pattern of ${\cal O}_{\pm}$. Even with an active charm, 
a quadratic divergent contribution needs to 
be subtracted in the parity-conserving sector
\ba
\widehat {\cal O}^{\rm PC}_{\pm}(\mu) & = & {\cal Z}_{\pm}(\mu) 
\Bigl[{\cal O}^{\rm PC}_{\pm} + \sum_j {b}^j_\pm {\cal O}_{\pm}^j +\nonumber\\
& + &  {b}_{\pm}^{\tau}{\cal Q}_\tau + 
\frac{{b}_{\pm}^s}{a^2} {\cal Q}_s \Bigr]\, ,
\ea
where 
\ba
{\cal Q}_s     & = & (m_u - m_c) \bar s  d \\
{\cal Q}_\tau  & = & (m_u - m_c) \bar s\, \sigma_{\mu\nu} F_{\mu\nu} d  \; .
\ea
and ${\cal O}^j_\pm$ are four-fermion operators with wrong chirality.
In this case only the flavour part of the GIM mechanism survives thanks 
to the explicit breaking of chiral symmetry \cite{Maiani:1986db}.
It is interesting to notice that power divergences can be mitigated with
twisted-mass Wilson fermions \cite{Pena:2002wj}.

At leading order in chiral perturbation theory, the 
matrix elements needed for the $\Delta I =1/2$ rule 
can be extracted from three-point correlation functions,
thus avoiding the infrared problem that affects the direct 
computation of the $K\rightarrow \pi\pi$ matrix elements on the lattice 
\cite{Bernard:wf}. No large cancellations among leading order terms 
are expected in the ratio $|A_0/A_2|$; an enhancement
should therefore be visible already at this order. As a result, a combined 
use of fermions 
with an exact chiral symmetry and chiral perturbation theory can be 
the starting point to attack the $\Delta I=1/2$ rule.

In the past years the RBC and the CP-PACS collaborations have studied the 
$\Delta I=1/2$ rule\footnote{See Ref.~\cite{Pekurovsky:1998jd} for an 
earlier attempt with staggered fermions.} with domain-wall fermions with a finite fifth dimension 
\cite{Blum:2001xb,Noaki:2001un}.  They have 
computed the $K\rightarrow \pi$ and $K\rightarrow 0$ matrix elements for 
the operators of the $\Delta S=1$ effective Hamiltonian in Eq.~(\ref{eq:nocharm})
and used LO chiral perturbation theory to recover 
the physical amplitudes. Albeit with large statistical and systematic
errors, both groups demonstrated that a controlled numerical signal can be 
obtained for these matrix elements. The systematics of these important results 
can be put under control by using fermions with an exact chiral symmetry, 
lighter quark masses, and 
by considering the effective Hamiltonian with a dynamical charm.
A numerical feasibility study with Neuberger's fermions is under way \cite{kpipi_club}.
It is also conceivable that the relevant low-energy constants of the 
weak chiral Lagrangian can be extracted by studying the weak interactions 
in the $\epsilon$-regime 
\cite{kpipi_club,Giusti:2002rx,Giusti:2002sm,Hernandez:2002ds}. 

For direct CP violation, both the ultraviolet and the infrared problems are 
more severe. An active charm does not mitigate the ultraviolet 
renormalization, and divergent power subtractions are necessary.
The cancellation between two large competing contributions 
from ${\cal Q}_6$ and ${\cal Q}_8$ renders $\varepsilon'/\varepsilon$ very sensitive 
to higher order corrections in chiral 
perturbation theory. Leading-order terms may not 
be sufficient to reach a reliable prediction in the Standard 
Model \cite{Bertolini:1998vd,Pallante:2001he}.\\ 

I warmly thank A. Vladikas and T. DeGrand for sending 
their results and for interesting discussions about 
their work. Many thanks to L. Lellouch for a critical reading 
of the manuscript and to the organizers for the very 
stimulating atmosphere of the conference. The participation
to the conference was supported in part by the European Community's 
Human Potential Programme under contract HPRN-CT-2000-00145, 
Hadrons/Lattice QCD.

%
%

\end{document}